\documentclass[twocolumn,showpacs,prb,10pt]{revtex4-1}
\usepackage[english]{babel}
\usepackage{amsmath,amssymb}
\usepackage{times}
\usepackage{epsfig}
\usepackage[colorlinks,linkcolor = blue,citecolor = blue]{hyperref}
\usepackage{color}
\begin{document}
\title{Spin hydrodynamics in the $S = 1/2$ anisotropic Heisenberg chain}
\author{J. Herbrych$^{1}$}
\author{R. Steinigeweg$^{1}$}
\author{P. Prelov\v{s}ek$^{1,2}$}
\affiliation{$^1$J. Stefan Institute, SI-1000 Ljubljana, Slovenia}
\affiliation{$^2$Faculty of Mathematics and Physics, University of Ljubljana, SI-1000 Ljubljana,
Slovenia}
\date{\today}


\begin{abstract}
We study the finite-temperature dynamical spin susceptibility of the one-dimensional (generalized) 
anisotropic Heisenberg model within the hydrodynamic regime of small wave vectors and frequencies. 
Numerical results are analyzed using the memory function formalism with the central quantity being 
the spin-current decay rate $\gamma(q,\omega)$. It is shown that in a generic nonintegrable model 
the decay rate is finite in the hydrodynamic limit, consistent with normal spin diffusion modes. On 
the other hand, in the gapless integrable model within the XY regime of anisotropy $\Delta < 1$ the behavior 
is anomalous with vanishing $\gamma(q,\omega=0) \propto |q|$, in agreement with dissipationless uniform 
transport. Furthermore, in the integrable system the finite-temperature $q = 0$ dynamical 
conductivity $\sigma(q=0,\omega)$ reveals besides the dissipationless component a regular part with 
vanishing $\sigma_{\text{reg}}(q=0,\omega \to 0) \to 0$.
\end{abstract}

\maketitle

\section{Introduction}
The $S = 1/2$ anisotropic Heisenberg model (AHM) on a one-dimensional (1D) chain is one of the 
prominent models describing the physics of strongly interacting fermions on a lattice. 
The model is also well realized in several novel materials, e.g., in the 1D 
Mott-Hubbard insulator the relevant low-energy degrees of freedom remain spin 
excitations, and the closest realizations correspond to the isotropic case $\Delta = 1$. The 
integrability of the model via the Bethe Ansatz has so far led to a number of exact results 
regarding the spin dynamics at $T=0$,\cite{shastry90} as well as some thermodynamic results at $T > 0$. \cite{takahashi72}
On the other hand, it has been also realized that just the integrability itself, of this particular 
model or more generally, can be the origin of the anomalous behavior of transport quantities and 
low-energy dynamics. \cite{cast95,zotos96,zotos01,meisner07}

A specific consequence of model integrability on a chain of $L$ sites is the existence of a 
macroscopic number of conserved local quantities and related operators $Q_n, n = 1, \dots, L$ 
commuting with the Hamiltonian $[Q_n,H] = 0$ and with each other $[Q_n,Q_m] = 0$. For the 1D AHM a 
nontrivial example is $Q_3 = J^E$ representing the energy current and leading directly to its 
non-decaying behavior\cite{zotos97} and within the linear response to infinite thermal conductivity $\kappa(T)$ at 
any $T \geq 0$, being also the explanation for the very large spin contribution to heat conductivity 
in spin-chain materials. \cite{hess07}

We concentrate in this paper rather on spin dynamics and response, as evidenced by the dynamical 
spin susceptibility $\chi(q,\omega)$, with the emphasis on the hydrodynamic regime of small 
$(q,\omega) \to 0$. The behavior is also expected to be anomalous since it has been shown that spin 
conductivity $\sigma(q=0,\omega)$ has a ballistic (dissipationless) contribution -- spin stiffness 
$D(T) > 0$ -- again at any $T \geq 0$ within the gapless XY regime $\Delta < 1$. \cite{meisner07} Although the corresponding 
spin current $J^z_0$ is not a conserved quantity, the ballistic component has been well established 
via several connections: (a) the spin current is closely related to many-body (MB) level dynamics 
$\epsilon_n(\phi)$ with respect to flux $\phi$ \cite{kohn64} and their independent-like character 
in an integrable MB model\cite{cast95,herbrych11}  in contrast to level
repulsion in a generic  nonintegrable system, (b) via the limited decay of correlation functions due the 
overlap with conserved quantities \cite{zotos97}, $\langle J^z_0 Q_3 \rangle \neq 0$, at least for a partly 
polarized system with nonzero magnetization, $s \neq 0$, (c) the overlap with a steady state 
solution\cite{prosen11} which -- unlike other local conservation laws -- is not orthogonal to the 
spin current in the $s=0$ sector. While the behavior is evidently different\cite{james09} but as 
well highly nontrivial in the gapped Ising-like regime $s=0, \Delta > 1$ where results seem to favor 
$D(T) = 0$, \cite{zotos99,herbrych11,prosen11} there seems to be 
most controversy on the strict vicinity of the isotropic point $\Delta \sim 1, s \sim 0$. 
\cite{meisner07, karrasch11,herbrych11} There is much less known about the $T > 0$ spin dynamics 
extended to 
finite wave vectors $q > 0$ as well as frequencies $\omega > 0$, which has been considered so far in 
finite systems using exact diagonalization\cite{shastry08,naef98} and quantum Monte 
Carlo\cite{brenig10,stary97} (QMC). While for a generic nonintegrable spin system the hydrodynamic 
regime should reveal diffusive behavior for $(q,\omega) \to 0$, in the gapless integrable AHM a singular 
approach at $q \to 0$ is expected. In the latter case one of the open questions is the possible 
coexistence of a ballistic and diffusive processes at $q \to 0$. \cite{takigawa96,thurber01} Clearly, 
the central challenge is the design of a proper phenomenological description of the $T > 0$ 
hydrodynamic response of the gapless
integrable as well as weakly perturbed nonintegrable spin system.

In this paper we analyze spin hydrodynamics in the integrable and nonintegrable AHM by the numerical 
calculation of $T > 0$ dynamical spin correlations on finite chains. 
Numerical data are used as the input to the memory-function representation 
where the crucial result is the spin-current decay rate $\gamma(q,\omega)$ at $T > 0$. The latter 
quantity should be constant $\gamma(q,\omega) = \gamma_0$ within the diffusion regime, as is shown to 
be for the nonintegrable case. For the integrable gapless AHM within the XY regime, however, our results 
indicate vanishing $\gamma(q \to 0,\omega = 0) = 0$ as well as $\gamma(q = 0,\omega \to 0) = 0$. 
This is well consistent with the dissipationless transport at $q = 0$ but puts also restrictions on 
possible proposed theoretical scenarios.

The paper is organized as follows: In Sec.~\ref{memory} we present the model, spin correlations and 
the memory-function formalism for the dynamical spin susceptibility. 
Section \ref{numerics} is devoted to the presentation of numerical methods. We focus also on 
the evaluation of the decay rate  $\gamma(q,\omega)$ from dynamical spin correlation functions. 
Results are presented in 
Sec.~\ref{nonintegrable} and Sec.~\ref{integrable}. First we investigate transport properties in a 
nonintegrable case with a next-nearest neighbor interaction. Next we repeat our 
investigation for high- and low-$T$ behavior of $q$ dependent $\gamma(q,\omega)$ in the case of the
integrable AHM. We also extend our analysis for the case of finite magnetization $s 
\ne 0$. Finally, we focus in Sec.~\ref{conductivity} on the low-$\omega$ behavior of the uniform 
dynamical conductivity.

\section{Memory-function analysis}
\label{memory}
In this paper we study the anisotropic Heisenberg (XXZ) model on a chain with $L$ sites and periodic 
boundary conditions (PBC),
\begin{equation}
H = J\sum_{i = 1}^L\left(S_i^xS_{i+1}^x+S_i^yS_{i+1}^y
+\Delta S_i^zS_{i+1}^z+\Delta_2S_i^zS_{i+2}^z\right) \, , \label{H}
\end{equation}
where $S_i^\alpha$ ($\alpha = x,y,z$) are spin $S = 1/2$ operators at site $i$ and $\Delta$ 
represents the anisotropy. We allow also for a next-nearest neighbor $zz$-interaction with $\Delta_2 
\neq 0$ breaking the integrability of the model. It should be reminded that the Hamiltonian 
(\ref{H}) can be mapped on a $t$-$V$-$W$ model of interacting spinless fermions with hopping $t = 
J/2$ and inter-site interactions $V = J\Delta$, $W = J\Delta_2$. In the fermionic representation, 
spin transport corresponds to charge transport. We further on use $\hbar = k_B = 1$ as well as $J = 
1$ as the unit of energy.

Our aim is to investigate the dynamical spin susceptibility $\chi(q,\omega)$ given by
\begin{equation}
\chi(q,\omega) = \imath\int\limits^\infty_0\text{d}t \, e^{\imath\omega t} 
\langle[S^z_{-q}(t),S^z_q]\rangle
\label{chiqom}
\end{equation}
with
\begin{equation}
S_q^z = \frac{1}{\sqrt{L}}\sum_ie^{\imath qi}S_i^z,
\end{equation}
where due to PBC $q = 2\pi k/L$ and $\langle\ldots\rangle$ denotes the thermodynamic average at 
temperature $T$. Note that the dynamical spin structure factor is related to the imaginary 
part of Eq.~(\ref{chiqom}) by $S(q,\omega) = \pi[1-\exp(-\omega/T)]\chi''(q,\omega)$.

While the results for $\chi(q,\omega)$ are obtained numerically for finite systems, we use for 
further analysis the memory-function (MF) 
formalism\cite{mori65,forster75,jung07} with the central quantity being 
the relaxation function
\begin{equation}
\Phi(q,\omega) = \frac{\chi(q,\omega)-\chi^0_q}{\omega} = (S^z_q|({\cal L}-\omega)^{-1}|S^z_q) \, ,
\label{rr2}
\end{equation}
where $\chi^0_q = \chi(q,0) = (S^z_q|S^z_q) $ is the static spin susceptibility, ${\cal L}$ is the 
Liouville operator, ${\cal L}A = [H,A]$, and within the MF formalism the scalar product between operators 
at $T > 0$ and corresponding $\beta = 1/T$ is defined by
\begin{equation}
(A|B) = \frac{1}{\beta}\int\limits_0^\beta \text{d}\tau\langle{A}^\dag B(\imath\tau)\rangle \, .
\end{equation}
Applying Mori-Zwanzig reduction\cite{mori65,forster75} we get
\begin{equation}
\Phi(q,\omega) = \frac{-\chi^0_q}{\omega+\imath q^{2}\tilde\sigma(q,\omega)/\chi^0_q} \, .
\end{equation}
Here, $\tilde\sigma(q,\omega)$ is the projected dynamical spin conductivity,
\begin{equation}
\tilde\sigma(q,\omega) = \imath (J_q^z|(\omega-Q{\cal L}Q)^{-1}|J_q^z) \, ,
\label{sigtilde}
\end{equation}
where the spin current is defined by the conservation law $\dot S_q ^z= i q J_q^ z$ (written at 
small momentum $q \ll 1$ ) and the orthogonal projection reads $Q = 1-|S^z_q)(S^z_q|/\chi^0_q$. The spin 
diffusion constant (if it exists) is equal to ${\cal D} = \lim_{q \to 0}\tilde\sigma(q,\omega = 
0)/\chi^0_q$. 

On the second level we can represent
\begin{equation}
\tilde\sigma(q,\omega) = \frac{\imath\varkappa_q^0}{\omega+M(q,\omega)} \, ,
\label{to}
\end{equation}
where $\varkappa^0_q = (\dot{S}^z_q|\dot{S}^z_q)/q^2 = (J^z_q|J^z_q)$, such that 
\begin{equation}
\Phi(q,\omega) = \frac{-\chi^0_q}{\omega-
\displaystyle{\frac{q^2\varkappa^0_q/\chi^0_q}{\omega+M(q,\omega)}}} \, .
\label{rr3}
\end{equation}

The memory function $M(q,\omega)$, and in particular its imaginary part $\gamma(q,\omega) = 
M''(q,\omega)$ representing the spin-current decay rate, is the central quantity of our analysis. In 
the generic case of normal transport it should be well behaved in the hydrodynamical regime 
$(q,\omega) \to 0$ with a finite value $\gamma_0 = \gamma(q \to 0,\omega \to 0)$. This is clearly 
not the case for the dissipationless (integrable) case where one can expect $\gamma(q \to 0,\omega = 
0) \to 0$.

On the other hand, $\chi_q^0$ and $\varkappa_q^0$ are both finite in the limit $q \to 0$. Moreover, 
both are even $q$ independent at high $T \gg 0$ ($\beta \to 0$), as can be easily shown via the
high-$T$ expansion,
\begin{eqnarray}
\chi_q^0 = \beta\left(1/4-s^2\right)+{\cal O}(\beta^2) \, ,\nonumber\\
\varkappa^0_q = \beta\frac{J^2}{2}\left(1/4-s^2\right)+{\cal O}(\beta^2) \, ,
\label{htchi0}
\end{eqnarray}
where $s = (N_\uparrow-N_\downarrow)/2L$ is the magnetization and $N_\uparrow$ ($N_\downarrow$) is 
number of up (down) spins.

It should be pointed out that, due to the projection $Q$ in Eq.~(\ref{sigtilde}), 
$\tilde\sigma(q,\omega)$ is not equal to the ``standard'' spin conductivity $\sigma(q,\omega)$,
\begin{equation}
\sigma (q,\omega) = \imath (J_q|(\omega-{\cal L})^{-1}|J_q) \, .
\end{equation}
Note that the latter quantity is directly related to the relaxation function $\sigma'(q,\omega) = 
\omega^2 \Phi''(q,\omega)/q^2$ and thus differs essentially from $\tilde\sigma'(q,\omega)$ for $q > 
0$, in particular in the regime $\omega \to 0$. On the other hand, for strictly $q = 0$ (in a finite 
system with PBC) both quantities should be equal, $\tilde\sigma(q=0,\omega) = \sigma(q=0,\omega)$, but 
in this case one has to calculate numerically $\sigma(0,\omega)$ and corresponding $M(q = 0,\omega)$ 
separately [not from $\Phi''(0,\omega)$] since $S_{q=0}^z$ is a conserved quantity.

\section{Numerical methods}
\label{numerics}
In this paper we present results obtained using different numerical methods:

\noindent (a) For high $T > 1$ studies we use the Microcanonical Lanczos 
Method\cite{prelovsek11,long03} (MCLM). The choice is motivated by the fact that in the 
thermodynamic limit the microcanonical ensemble should yield the same results as the canonical one. 
For a large enough system $L$ and temperature $T$, dynamical autocorrelations can be evaluated with 
respect to a single wave function $|\Psi\rangle$ characterized by the energy uncertainty
\begin{equation}
\delta\epsilon = \sqrt{(\langle \Psi|(H-\lambda)^2|\Psi \rangle)} \, ,
\end{equation}
where the parameter $\lambda = \langle H \rangle$ determines the temperature 
for which $|\Psi\rangle$ is a relevant representative. Such 
$|\Psi\rangle$ can be generated via a first Lanczos procedure using $(H-\lambda)^2$ instead of $H$. 
The dynamical correlation functions are then calculated using the standard Lanczos method, where the 
modified wave function $\tilde\Psi = S^z_q|\Psi\rangle$ is the starting point for the second Lanczos 
iteration. Reachable finite-size systems, $L = 28$ for magnetization $s = 0$ and $L = 36$ for $s = 
1/4$, have very high density of states for high $T$, hence statistical fluctuations are effectively 
smoothed out. This is in contrast to low-$T$ properties, dominated by a small number of low-lying MB 
states, therefore a different numerical method should be applied.

\noindent (b) For low $T<1$ we choose the Finite-Temperature Lanczos 
Method\cite{prelovsek94,prelovsek11} (FTLM) which is reliable for $T > T_{\text{fs}}$ where the
finite size temperature is typically $T_{\text{fs}} \sim 0.5$ in the AHM at available chain lengths $L $. 
To reduce  statistical fluctuations at low $T$, one can introduce additional sampling over initial random 
vectors and increase the number of Lanczos steps. Due to the memory requirement and the CPU time, 
reachable system sizes are $L = 26$ for $s = 0$ and $L = 32$ for $s = 1/4$.

Within the framework of linear response theory we numerically calculate the imaginary part of the 
dynamical susceptibility $\chi''(q,\omega)$ and for $q=0$ the spin dynamical conductivity 
$\sigma'(0,\omega)$. The MF $\gamma(q,\omega)$ can be calculated directly from Eq.~(\ref{rr3}), 
\begin{equation}
\gamma(q,\omega) =
\Im\left(\frac{q^{2}\varkappa^0_q\Phi(q,\omega)}{\chi^0_q[\chi^0_q+\omega\Phi(q,\omega)]}\right) 
\, . \label{gama}
\end{equation}
However, the denominator of this equation is highly 
influenced by the Kramers-Kronig transformation, e.g., by a small imaginary process $\epsilon$ added 
in the numerical realization of the Kramers-Kronig relation, 
\begin{equation}
\Phi(q,\omega) = -\frac{1}{\pi}\int\limits_{-\infty}^\infty
\text{d}\omega'\frac{\Phi''(q,\omega')}{\omega-\omega'+\imath\epsilon} \, .
\end{equation}
To resolve this problem it is better to perform the Hilbert transform on $\chi''(q,\omega)$ or 
$\sigma'(0,\omega)$ and use one of the following expressions for the MF,
\begin{eqnarray}
\gamma(q = 0,\omega) = \frac{\varkappa^0_0\sigma'(q = 0,\omega)}{|\sigma(q = 0,\omega)|^2} \, ,
\nonumber\\
\gamma(q,\omega) = \frac{q^2\varkappa^0_q\chi''(q,\omega)}{\omega|\chi(q,\omega)|^2} \, ,
\label{mq}
\end{eqnarray}
which comes from simple evaluation of Eq.~(\ref{to}) and Eq.~(\ref{rr3}) respectively. Finally, we 
can calculate $\varkappa_q^0$ using one of the sum rules
\begin{equation}
\varkappa^0_q = \int\frac{\text{d}\omega}{\pi}\,\omega\frac{\chi''(q,\omega)}{q^2} \, , \quad
\varkappa^0_0 = \int\frac{\text{d}\omega}{\pi}\,\sigma'(q = 0,\omega) \, ,
\end{equation}
where the integration goes over the whole, numerically obtained MB spectrum, i.e., $\omega_{\text{span}} 
\sim LJ $. It is also worth mentioning that $\chi^0_q$ and $\varkappa^0_q$ are weakly $q$ dependent 
in the hydrodynamic limit, as evident for $\beta \to 0$, cf.~Eq.~(\ref{htchi0}).
\begin{figure}[t]
\includegraphics[width=0.85\columnwidth]{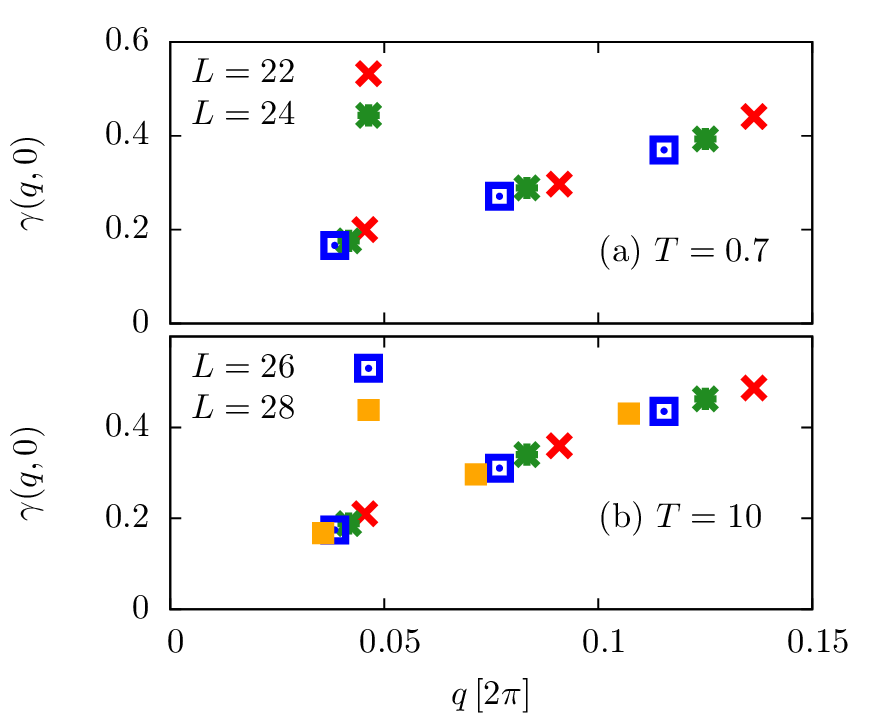}
\caption{(Color online) $\gamma(q,\omega=0)$ vs.~$q$ (in real units) for $\Delta = 0.5$, $s = 0$,
as calculated for a integrable system of $L = 22,24,26$, and $28$ sites.}
\label{Lsc}
\end{figure}

As a final remark of this section we show in Fig.~\ref{Lsc} result for the decay rate
$\gamma(q,\omega=0)$, evaluated for different system sizes $L=22-28$. We choose here $\Delta=0.5$ and 
$\Delta_2=0$ case, but this behavior is generic for ``conducting'' regime. It is evident that $\omega=0$
limit of decay rate revel only systematic size ($q$) dependence, without sizeable statistical
error or deviations.
\section{Nonintegrable case}
\label{nonintegrable}
\begin{figure}[t!]
\includegraphics[width=0.85\columnwidth]{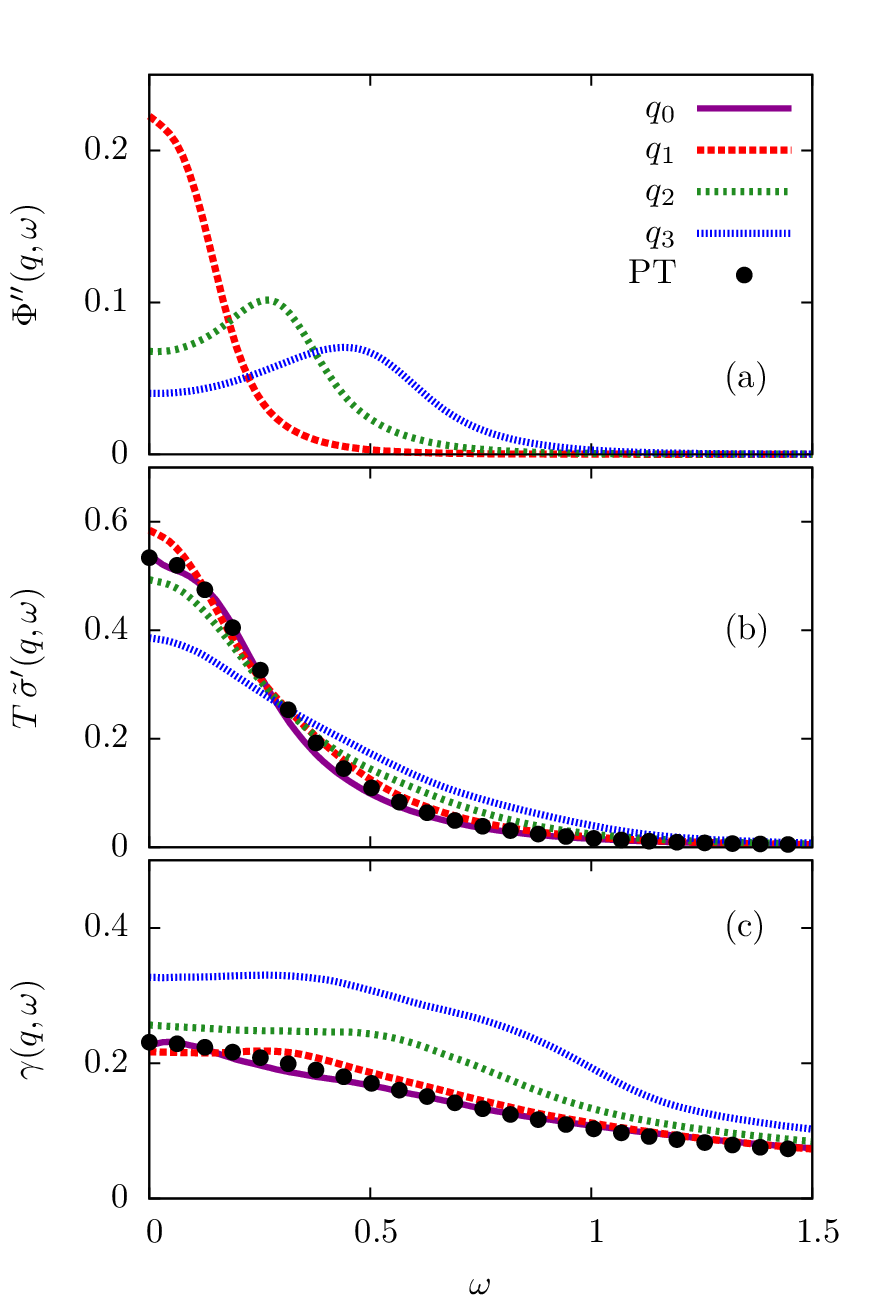}
\caption{(Color online) High-$T$ (a) spin relaxation function $\Phi''(q,\omega)$, (b) projected 
conductivity $\tilde\sigma(q,\omega)$ and (c) spin-current decay rate $\gamma(q,\omega)$ for a nonintegrable AHM 
at $s=0$, $T = 10$, $\Delta = 0.5$, $\Delta_2 = 0.6$, as calculated for a 
system of $L = 28$ sites. With dots we present the comparison with a $q=0$ perturbation theory.}
\label{NI}
\end{figure}

First, let us address the question of spin transport in the generic ``normal'' case, i.e., 
nonintegrable case as introduced by $\Delta_2 \ne 0$. In Fig.~\ref{NI} we present characteristic 
numerical results within the XY regime $\Delta = 0.5$ at high $T$, as obtained via the MCLM method 
for a chain of $L = 28$ sites and different allowed smallest $q_n = n(2\pi/L), n = 0,\dots,3$.

Shown are the dynamical relaxation function $\Phi''(q,\omega) = \chi''(q,\omega)/\omega$, and the 
extracted projected dynamical conductivity $\tilde\sigma'(q,\omega)$, Eq.~(\ref{to}), as well as the 
decay rate $\gamma(q,\omega)$, Eq.~(\ref{mq}). Results obtained here for a substantial integrability 
breaking term $\Delta_2 = 0.6$ can be easily interpreted with normal diffusion behavior. We note 
that the projected conductivity $\tilde\sigma(q,\omega)$ reveals a nearly $q$ independent Lorentzian 
form, which is then reflected in $\gamma(q,\omega)$, being effectively constant in both $\omega$ and $q$. 
In particular, it is evident that the $q=0$ result $\gamma(q=0,\omega)$ obtained directly from 
$\sigma(q=0,\omega)$ is essentially the same as $\gamma(q_1,\omega)$ extracted from 
$\Phi''(q_1,\omega)$. Deviations for $q > q_1$ are understandable since for our restricted chain 
lengths the latter $q$'s are actually not very small.
\begin{figure}[t!]
\includegraphics[width=0.85\columnwidth]{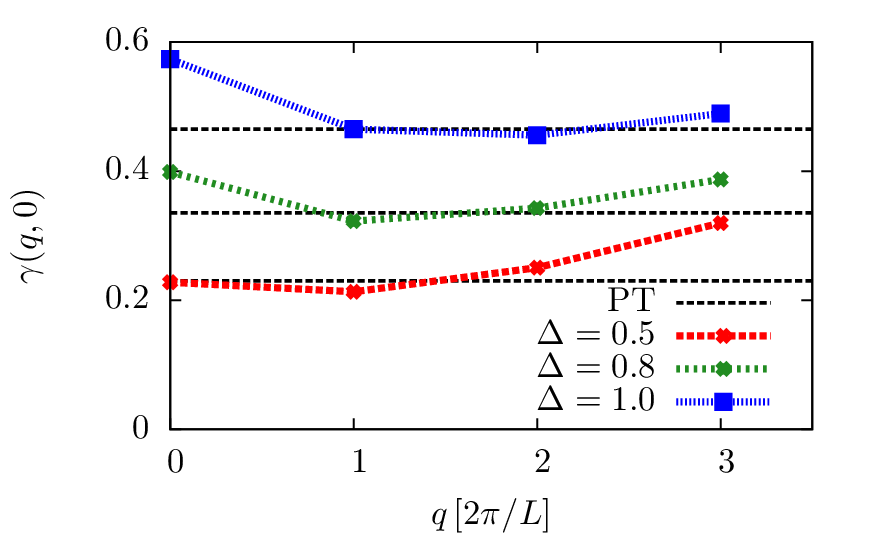}
\caption{(Color online) Decay rate $\gamma(q,\omega=0)$ vs. momentum $q$ for the case as
in Fig.~\ref{NI} with $\Delta_2 = 0.6$, but for different $\Delta = 0.5, 0.8, 1.0$, evaluated for $L 
= 28$. Horizontal lines represent $q=0$ results from the perturbation theory.}
\label{NQ}
\end{figure}

As a final numerical result of this section we present in Fig.~\ref{NQ} the finite $q$-scaling of 
the d.c.~rate $\gamma(q,\omega=0)$ for three different values of the anisotropy $\Delta = 
0.5$, $0.8$, $1$ in the case of non-integrability $\Delta_2 = 0.6$ of the model (\ref{H}). As 
already evident from Fig.~\ref{NI}, the variation with $q$ is modest and $\gamma(q,\omega=0)$ is finite in 
the limit $q \to 0$. Furthermore, the value agrees very well with a perturbation theory at $q=0$ on 
the basis of Ref.~\onlinecite{steinigeweg2011}, on which we comment in more detail for the remainder 
of this section.

To this end, it is convenient to turn to the time domain for the moment and to consider an 
integro-differential equation of the form
\begin{equation}
\partial_t \sigma'(q=0,t) = -\int\limits_0^t \text{d}\tau \, K(t-\tau) \sigma'(q=0,\tau) \label{NZ}
\end{equation}
describing the decay of the dynamical conductivity $\sigma'(q=0,t)$ in time and involving a 
time-convolution with a memory kernel $K(t)$. Such an equation, and particularly the memory kernel, 
can be obtained by an application of the Nakajima-Zwanzig (NZ) projection operator technique. \cite{breuer07} For 
the application of NZ an unperturbed Hamiltonian $H_0$ has to be chosen, where a natural choice is 
given by the XX model, i.e., $H_0 = H(\Delta=\Delta_2=0)$. Then the interaction $V = H-H_0$ may play 
the role of a small perturbation and NZ allows to obtain the memory kernel $K(t)$ as a lowest-order 
truncation of a systematic series expansion (in even powers of the perturbation strength involving 
$\Delta$ and $\Delta_2$). The somewhat lengthy and subtle calculation of the lowest-order truncation 
$K_2(t)$ has already been undertaken and is given in Eq.~(30) of Ref.~\onlinecite{steinigeweg2011} 
by $K_2(t) = \partial_t R^{1,1}_2(t)$. While the final expression for $K_2(t)$ still needs to be 
evaluated numerically, the evaluation can be done for several thousands sites, e.g., $L = 2000$ as 
chosen here. Thus, numerically integrating the above Eq.~(\ref{NZ}) by the use of $K(t) \approx 
K_2(t)$ for $L=2000$ eventually leads to a lowest-order prediction in the thermodynamic limit for 
$\sigma'(q=0,t)$, and after Fourier transforming, for $\sigma'(q=0,\omega)$ as well. The obvious 
agreement in Fig.~\ref{NI} also includes small deviations from a strict Lorentzian, which is a 
non-Markovian effect since the memory kernel $K_2(t)$ cannot be considered as a mere $\delta$ 
function on the characteristic scale set by the decay time of $\sigma'(q=0,t)$. Notably, $K_2(t)$ 
directly yields a lowest-order prediction on the frequency-dependent decay rate 
$\gamma(q=0,\omega)$ as well, via the relation
\begin{equation}
\gamma(q=0,\omega) \approx \int\limits_0^\infty \! \text{d}t \, \cos(\omega t) K_2(t) \, ,
\end{equation}
with an as convincing agreement in Figs.~\ref{NI} and \ref{NQ}. It is further worth mentioning that 
the success of NZ relies not only on the limit of small perturbations but also crucially on the 
non-integrability, 
allowing to assume vanishing higher-order contributions (equivalent to neglecting the coupling to 
other observables in the NZ equation). In the case of integrability, or in the limit of strong 
perturbations, the incorporation of higher order contributions is indispensable and also other 
variants of projection operator techniques may become convenient \cite{steinigeweg2011}, not 
containing a memory kernel at all.
\begin{figure}[t]
\includegraphics[width=0.85\columnwidth]{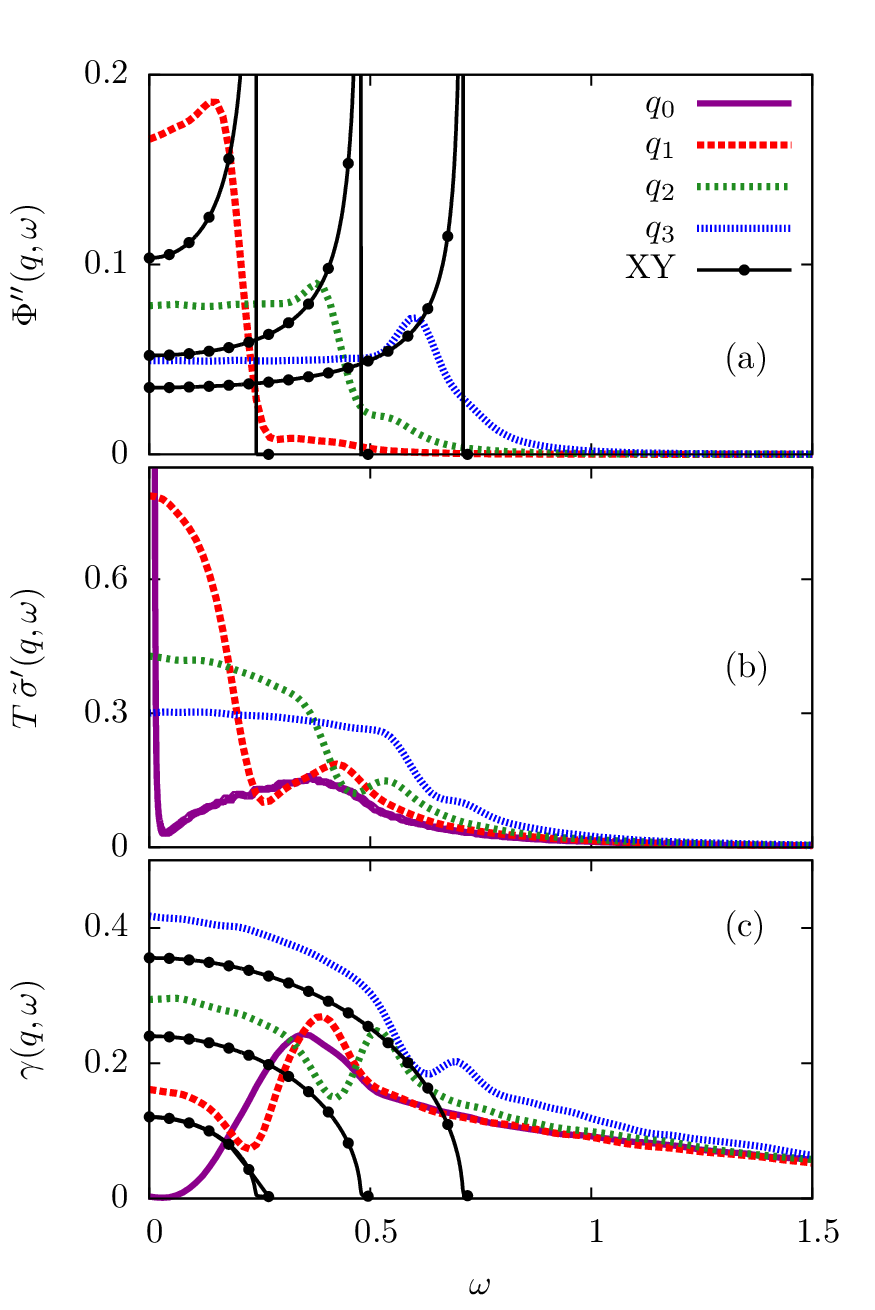}
\caption{(Color online) High-$T$ (a) $\Phi''(q,\omega)$, (b) $\tilde\sigma(q,\omega)$ and (c) 
$\gamma(q,\omega)$ for the integrable case at $s = 0$, $\Delta = 0.5$, evaluated for $L = 28$ and $T 
= 10$. Points represent the analytical $\Delta=0$ results for the same parameters and $q$.}
\label{I}
\end{figure}
\begin{figure}[t]
\includegraphics[width=0.85\columnwidth]{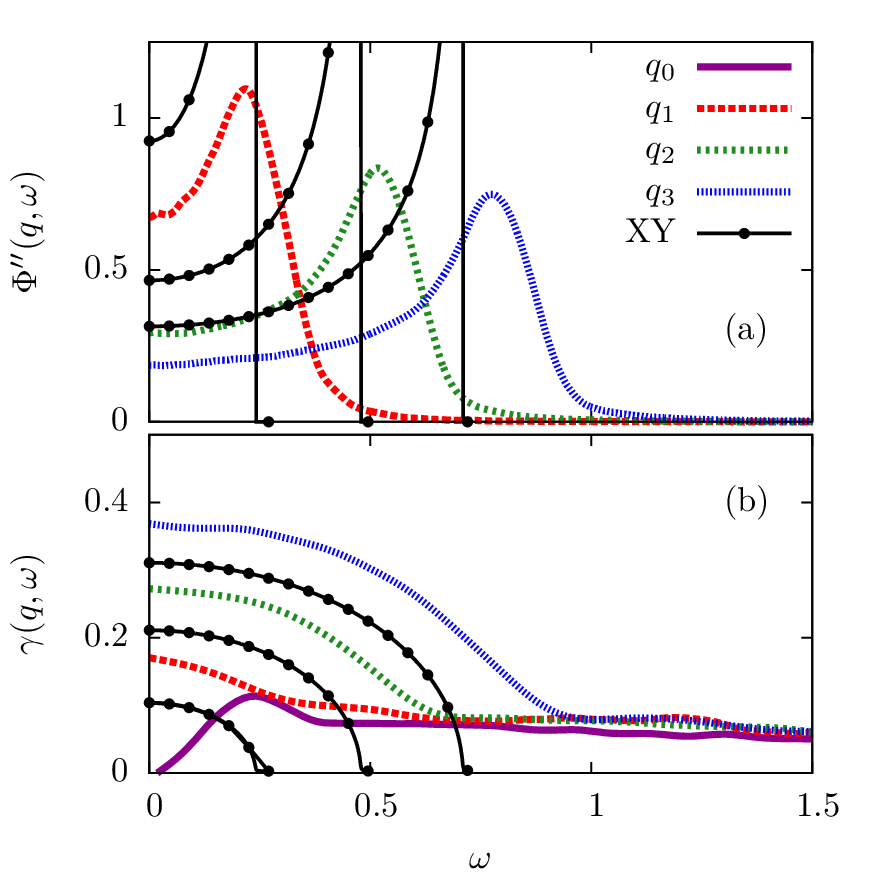}
\caption{(Color online) Low-$T$ (a) $\Phi''(q,\omega)$ and (b) $\gamma(q,\omega)$ for the integrable 
case as calculated for $L=26$, $s = 0$, $\Delta = 0.5$, and $T=0.7$. Points represent the analytical $\Delta=0$ results 
for the same parameters and~$q$.}
\label{IFT}
\end{figure}
\section{Integrable model}
\label{integrable}
In the following we analyze the spin dynamics within the integrable AHM (XXZ) model as realized for 
$\Delta_2 = 0$. We concentrate here on the gapless (``conducting'') regime, i.e.~on the XY regime $\Delta < 
1$ at zero magnetization $s = 0$, but extending also to $\Delta \leq 1$ for $s \neq 0$. Note that a 
different behavior can emerge in the  ``insulating'' (spin-gap) phase at $\Delta \geq 1$ and $s = 
0$ (even at $T > 0$).

\subsection{XX model}
In the particular case $\Delta = 0$ the AHM (i.e. the XX model) can be mapped on the tight-binding 1D model of 
noninteracting spinless fermions. In this case the calculation of the dynamical susceptibility, 
Eq.~(\ref{chiqom}), reduced to the well known Lindhard formula and for $q \to 0$ the relaxation 
function $\Phi''(q,\omega)$ can be evaluated analytically for arbitrary $T > 0$,
\begin{equation}
\Phi''(q,\omega) = \frac{\beta}{4\cosh^2\left(\beta\xi/2q\right)\xi} \, , \qquad
\xi = \sqrt{q^2-\omega^2} \, .
\label{pe}
\end{equation}
Using the above expression and relation (\ref{gama}) one can express directly the decay 
rate $\gamma(q,\omega)$, and in particular
\begin{equation}
\gamma^{XX}(q,\omega=0) = \frac{\varkappa^0_q\beta}{[2\chi^0_q\cosh(\beta/2)]^2}|q| \, .
\label{xy}
\end{equation}
Since the prefactor at any $T > 0$ is a constant at $q \to 0$, we note a characteristic 
linear variation $\gamma^{XX} \propto |q|$ and its (nonanalytic) vanishing for $q \to 0$. As we see 
in the following, the XX model can serve as the guideline and explanation for the observed behavior 
for more general $\Delta >0$.

\subsection{XXZ model}
Next let us focus on the case of a more general AHM with $\Delta>0$ (XXZ model) which corresponds, via the 
Jordan-Wigner transformation, to a model of interacting spinless fermions and analytical results are 
not available for $T>0$ and $q>0$. We are therefore restricted to the numerical calculation of 
$\Phi(q,\omega)$. As described in Sec.~\ref{numerics}, two approaches are used: (a) the MCLM for 
high enough $T > 1$ and (b) the FTLM for lower $T$. 

In Fig.~\ref{I} we first present the high-$T$ results for $\Delta = 0.5$ and $s=0$. Again we show 
first $\Phi''(q,\omega)$, then the extracted $\tilde\sigma'(q,\omega)$, and finally 
$\gamma(q,\omega)$. The difference to a nonintegrable case is quite evident. Both $\tilde 
\sigma'(q,\omega)$ as well as $\gamma(q,\omega)$ are strongly $q$ dependent, in particular at low 
$\omega$, most pronounced for $q=0$. Here, $\tilde\sigma'(q=0, \omega)= \sigma'(q=0, \omega)$ 
exhibits a dissipationless component $2 \pi D \delta(\omega)$ reflected in the vanishing $\gamma(q=0, 
\omega \to0)\to 0$. 

To stress the correspondence of the results to the XX case we present in Figs.~\ref{I} 
and ~\ref{IFT} results for $\Phi''(q,\omega)$ and $\gamma(q,\omega)$ in a direct comparison with the 
analytical $\Delta=0$ expressions Eq.~(\ref{pe}) with the same parameters, same 
$q$, as well as $T=10$ and $T=0.7$, respectively. Similarities and differences to the XX case are 
quite visible: (a) singularities and cutoffs in $\Phi''(q,\omega)$ at $\omega_c(q) = c q$ persist at 
$\Delta >0 $ as maxima diminishing with increasing $\Delta$, (b) the qualitative behavior of 
$\gamma(q,\omega)$ is quite similar for the $\Delta=0$ and $\Delta=0.5$ cases for lower $\omega < 
\omega_c$, (c) for $\Delta=0$ decay rate vanishes above the threshold, i.e., $\gamma[q, \omega> 
\omega_c(q)] =0$, for $\Delta>0$ on the other hand $\gamma[q,\omega>\omega_c(q)]$ becomes $q$ 
independent but nonzero in a broad $\omega$ range.

\begin{figure}[t]
\includegraphics[width=0.85\columnwidth]{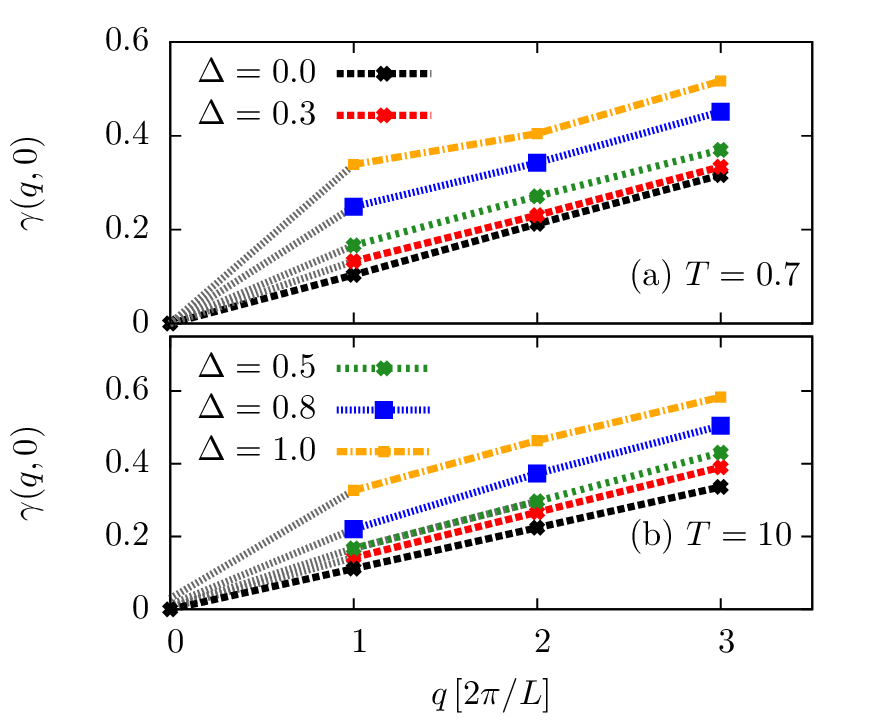}
\caption{(Color online) $\gamma(q,\omega=0)$ vs.~$q$ for $s=0$ and different 
$\Delta \leq 1$, (a) low $T=0.7$ and $L=26$, (b) $T \gg 1$ and $L=28$. Gray-dashed lines are guides 
to the eye.}
\label{Q}
\end{figure}

Next we focus on the finite $q$-scaling of $\gamma(q,\omega)$. In Fig.~\ref{Q} we present $\gamma(q,\omega=0)$
for the zero-magnetization $s=0$ as a function of $q$ for various values 
of the anisotropy ($\Delta=0,0.3,0.5,0.8,1$) and for two temperatures $T=0.7$ and $T=10$. 
As we mention before, the $q=0$ and $q \ne 0$ results are calculated from different (numerically 
obtained) quantities, dynamical conductivity $\sigma'(q=0,\omega)$ and spin susceptibility 
$\chi''(q,\omega)$, respectively. It is indicative that $\gamma(q,\omega=0)$ within the XY regime 
is linear function of $q$. Also our results show vanishing $\gamma(q \to 0,\omega=0) \to 
0$ for $\Delta < 1$. This type of behavior is well consistent with dissipationless (ballistic) 
transport for zero wave vector $q=0$. It is, however, also visible that the isotropic case $\Delta=1$ 
does not follow the simple $\gamma(q,\omega=0) \propto q$ behavior (similar deviations can be observed also 
for $\Delta  =0.8$ and low $T=0.7$ which could be also due to finite-size or FTLM numerical 
uncertainty). As discussed later the conclusion could be that $\gamma(q \to 0,\omega=0) = \gamma_0^0$ 
remains finite  \cite{sirker09} or  even more that the scaling $\gamma(q,\omega=0) \propto q^\alpha$ 
is different  \cite{znidaric11}.

\begin{figure}[t]
\includegraphics[width=0.85\columnwidth]{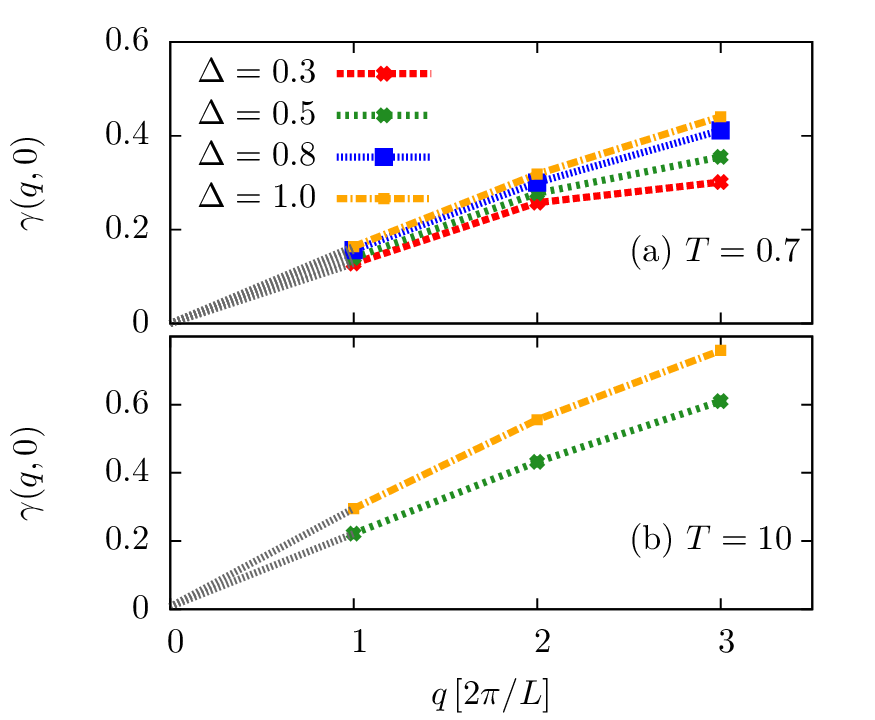}
\caption{(Color online) $\gamma(q,\omega=0)$ vs.~$q$ for $s=1/4$ and different 
$\Delta \leq 1$, (a) low $T=0.7$ and $L=32$, (b) $T \gg 1$ and $L=36$. Gray-dashed lines are 
guides to the eye.}
\label{QQF}
\end{figure}

We also repeat the same analysis for the case of finite magnetization $s \neq 0$ . 
Figure \ref{QQF} depicts the  $q$-scaling for $s=1/4$. Since we are away from the 
singular  isotropic point $s=0, \Delta=1$ and $D(T)>0$ for all $\Delta$, \cite{zotos97}
here the $\Delta=1$ case does not show any deviations from a general rule.
\begin{figure}[t]
\includegraphics[width=0.85\columnwidth]{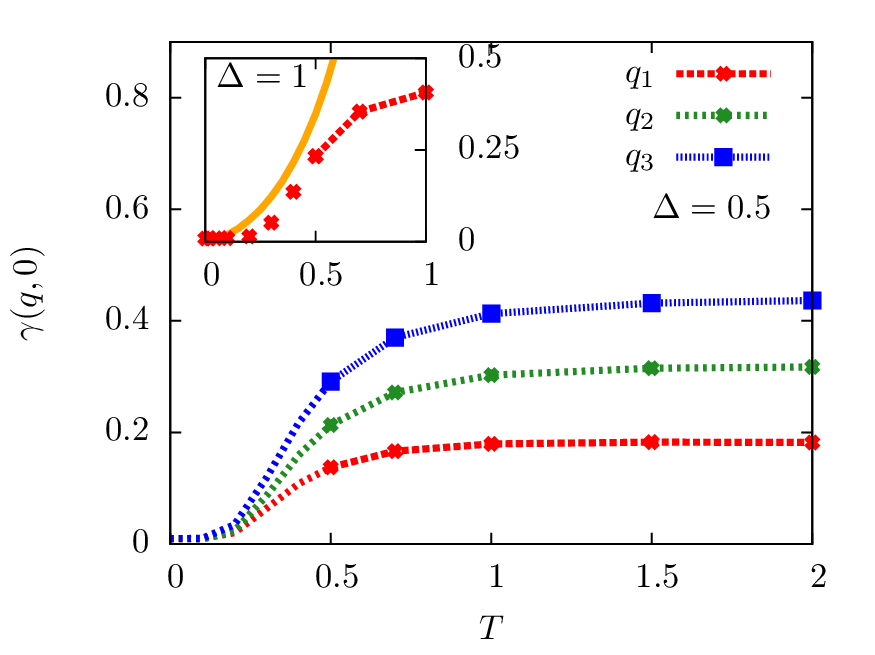}
\caption{(Color online) $\gamma(q,\omega=0)$ vs.~$T$ for $\Delta = 0.5$, $s = 0$, as calculated for a 
system of $L = 26$ sites. Inset: For $\Delta = 1$ the comparison between $\gamma(q_1,\omega=0)$ [dashed, red curve] 
and the bosonization result in Eq.~(\ref{gasi}) [solid, orange curve]. }
\label{TQ}
\end{figure}

Finally let us focus on the temperature dependence of $\gamma(q,\omega)$. As mentioned, the FTLM 
is reliable for $T > T_{\text{fs}} \sim 0.5$ while presented data for $T<0.5$ are only estimates. 
Figure~\ref{TQ} depicts results for $\Delta=0.5$, characteristic results for the 
XY regime. Several conclusions can be drawn 
directly from the obtained results: (a) For $T \geq 1$ the decay rate is effectively
$T$ independent, (b) In the regime $T<1$ but $T>T_{\text{fs}}$ where FTLM is reliable, 
it appears that $\gamma(q,\omega=0) \propto |q|$ for all $T$ but the $T$-dependence reveals
the vanishing decay rate for $T \to 0$ as expected for low $T$ where the model becomes
the one of free quasiparticles within the Luttinger 1D liquid.

In the inset of Fig.~\ref{TQ} we show a comparison with the bosonization result $\tilde\gamma$ in
Ref.~\onlinecite{sirker09} for $\Delta=1$ and $q=0$,
\begin{equation}
\tilde\gamma(T)=\pi g^2T,
\label{gasi}
\end{equation}
where the running coupling constant $g$ is determined by the equation
\begin{equation}
\frac{1}{g}+\frac{\ln g}{2}=\ln\left(\sqrt{\frac{\pi}{2}}\frac{e^{0.8272}}{T}\right) \, .
\end{equation}

We point out that our result, for the smallest possible $q_1=2\pi/L$ at $L=26$, is smaller in 
the entire $T>T_{\text{fs}}$ regime. It should be also noted $\gamma(q_1,\omega=0)$ is clearly an upper 
bound to the desired $\gamma(q \to 0,\omega=0)$. 

\section{Uniform dynamical spin conductivity}
\label{conductivity}
It is by now well accepted that the transport in an integrable 1D model can be dissipationless at 
any $T>0$ in the ``conducting'' regime. \cite{zotos97,meisner07}
For the uniform spin current, given explicitly as
\begin{equation}
J_0^z = J\sum_r \left(S^x_r S^y_{r+1} - S^y_r S^x_{r+1}\right) \, ,
\end{equation}
one can express the conductivity for $q = 0$ [$\sigma'(0,\omega) = \sigma'(\omega)$] at $T > 0$ as 
\begin{equation}
\sigma'(\omega) = 2\pi D\delta(\omega)+\sigma'_{\text{reg}}(\omega) \, ,
\end{equation}
where the regular part $\sigma'_{\text{reg}}(\omega)$ can be expressed in terms of eigenstates 
$|n\rangle$ and eigenenergies $\epsilon_n$,
\begin{equation}
\sigma'_{\text{reg}}(\omega) =
\frac{\pi}{L}\frac{1-e^{-\beta\omega}}{\omega}\sum_{\epsilon_n\ne\epsilon_m}p_n 
|\langle n|J_0^z |m\rangle|^2\delta(\epsilon_n-\epsilon_m-\omega) \, ,
\label{sr}
\end{equation}
while the dissipationless component with the Drude weight (spin stiffness) $D$ is related to 
matrix elements between degenerated states,
\begin{equation}
D = \frac{\beta}{L}\sum_{\epsilon_n = \epsilon_m}p_n|\langle n|J_0^z |m\rangle|^2 \, ,
\label{st}
\end{equation}
where $p_n = e^{-\beta\epsilon_n}/Z$ are corresponding Boltzmann factors and $Z$ is the partition 
function.
\begin{figure}[t]
\includegraphics[width=0.85\columnwidth]{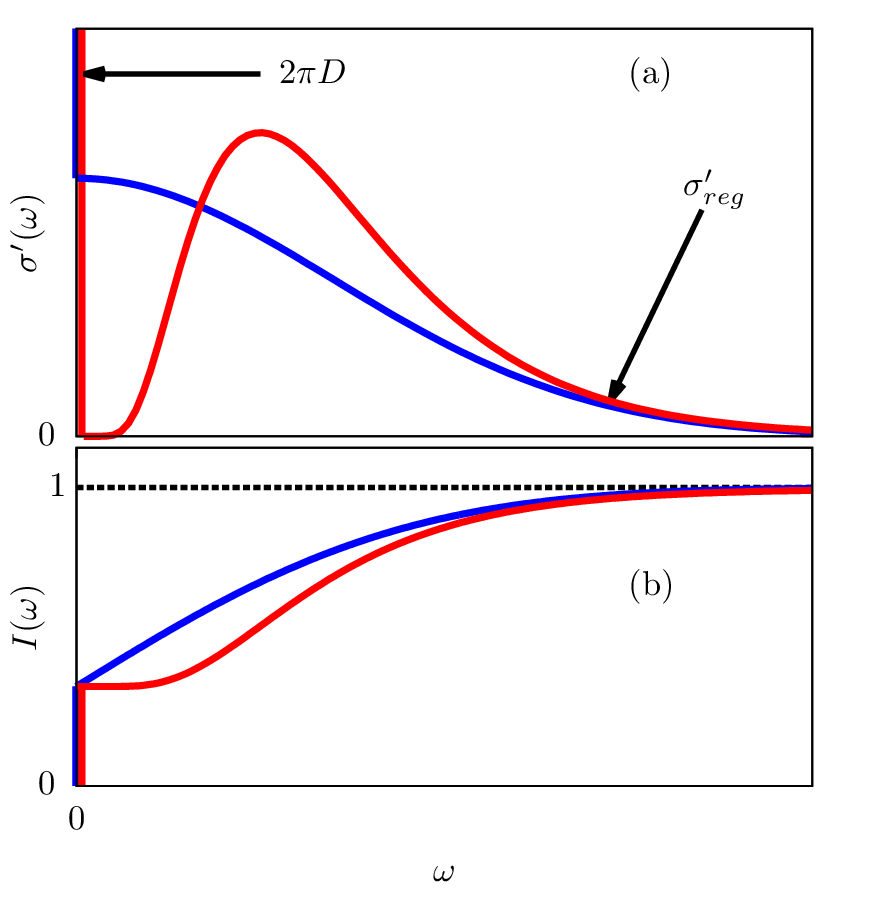}
\caption{(Color online) Sketch of two possible scenarios for (a) $\sigma'(\omega)$ and (b) the 
normalized integrated spectrum $I(\omega)$.}
\label{SCHE}
\end{figure}

While there are several analytical as well as numerical results supporting $D(T>0) >0$ in the XY 
regime $s=0, \Delta<1$ of the integrable AHM, \cite{zotos96,zotos99,meisner07,herbrych11,prosen11} 
there is still controversy on the behavior of the regular 
part $\sigma_{reg}'(\omega)$ in the same regime. The upper part of Fig.~\ref{SCHE} shows two possible 
scenarios for $\sigma'(\omega)$ differing essentially in the behavior at $\omega \to 0$. Note that 
both scenarios differ also in the integrated normalized spectra
\begin{equation}
I(\omega) = \frac{1}{\pi\varkappa^0_0}\int\limits_{-\omega}^\omega \text{d}\omega'\sigma'(\omega') 
\, ,
\end{equation}
also presented in Fig.~\ref{SCHE}, which are much more reliable (monotonically increasing function) 
when numerically dealing with finite-system results. Some applications of the MF
formalism combined with a coupling to conserved quantities and perturbative scattering of 
nonconserved quantities (similar to the NZ approach in this paper) indicate on 
$\sigma_{\text{reg}}'(0) \ne 0$. \cite{sirker09,rosch00} We will in the following present both an 
analytical argument as well as numerical results leading to the conclusion that this is not the case.

As the basis of the reasoning we follow Ref.~\onlinecite{shastry02} for the integrable 1D model, 
characterized by a macroscopic number of local conserved quantities $Q_n, n=1,\ldots,L$. Let us 
consider a perturbed Hamiltonian by a fictitious magnetic flux, \cite{shastry90,kohn64} which 
modifies the exchange (hopping) term in Eq.~(\ref{H}) by the Peierls phase factor $J \to 
J\exp(\imath\phi)$. It can be directly verified that the perturbed Hamiltonian is still 
characterized by the same conserved quantities, i.e., $[H(\phi),Q_n(\phi)]=0$. In 
particular, we employ here only $Q_3(\phi)=J_E(\phi)$ representing within the integrable AHM the 
conserved  energy current at $q=0$. The Taylor expansion in $\phi$ of $H(\phi)$ and $J^E(\phi)$ 
leads to an exact relation,
\begin{equation}
[H(\phi),W(\phi)]+[J_0^z(\phi),J^E(\phi)] = 0 \, ,
\end{equation}
where $W(\phi) = \partial J^E(\phi)/\partial\phi$ and $J_0^z(\phi) = \partial H(\phi)/\partial\phi$. 
Evaluating the matrix element of the above equation between eigenstates $|n\rangle$ and $|m\rangle$, 
we find
\begin{equation}
\langle n|J_0^z(\phi)|m\rangle = \langle n|W(\phi)|m\rangle\, 
\frac{\epsilon_n(\phi)-\epsilon_m(\phi)}{\jmath^E_n(\phi)-\jmath^E_m(\phi)} \, ,
\label{lc}
\end{equation}
where $\jmath^E_n$ are eigenvalues of $J^E$ and $\epsilon_n(\phi)-\epsilon_m(\phi) = -\omega$. It 
seems to be plausible \cite{shastry02,zotos96} that $\jmath^E_n$ do not have the same crossing 
points as $\epsilon_n$, such that the denominator in Eq.~(\ref{lc}) remains finite at $\omega \to 
0$. Taking this into account as well as Eq.~(\ref{sr}) we see that $\langle n|J_0^z(\phi)|m\rangle 
\propto \omega$ as $\omega \to 0$ and finally
\begin{equation}
\lim\limits_{\omega \to 0}\sigma'_{\text{reg}}(\omega) \propto \omega^2 \, . \label{regu}
\end{equation}
This scenario is depicted in Fig.~\ref{SCHE}. This behavior of $\sigma'(\omega)$ at low frequency 
is also clearly visible in the normalized integrated dynamical conductivity $I(\omega)$, see lower 
part of Fig.~\ref{SCHE}. Such a behavior has been already observed in 
Ref~\onlinecite{zotos96,shastry08,herbrych11,rigol08}. 
\begin{figure}[t]
\includegraphics[width=0.85\columnwidth]{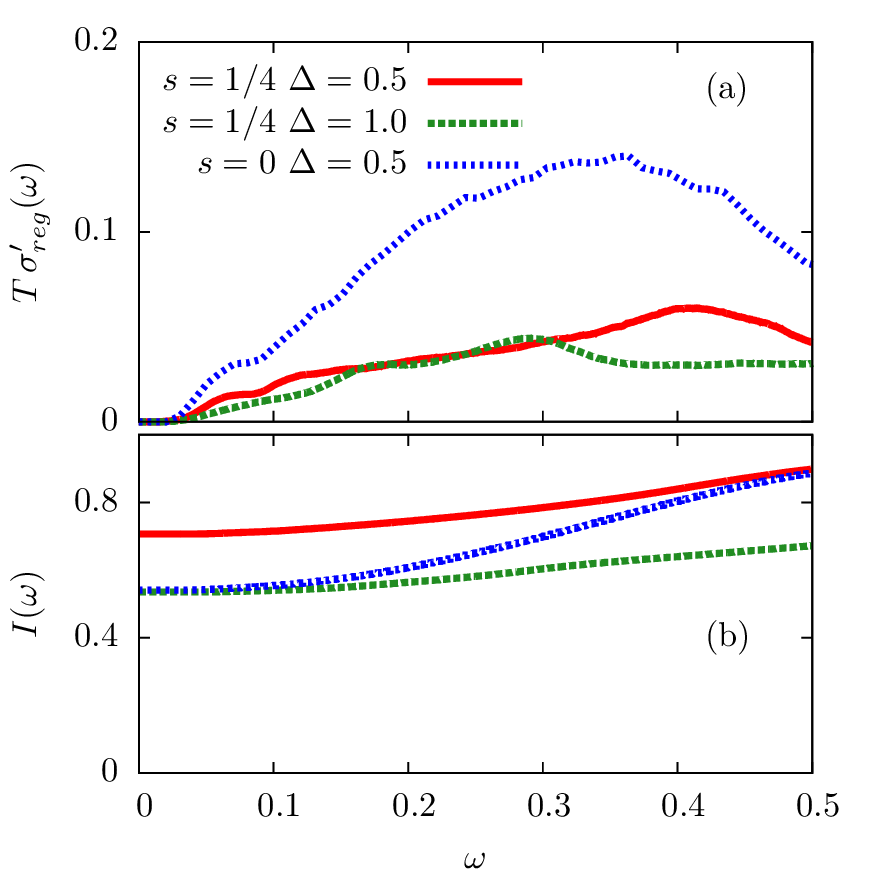}
\caption{(Color online) Low-$\omega$ dependence of (a) $\sigma'_{\text{reg}}(\omega)$ and (b)
normalized $I(\omega)$.}
\label{SI}
\end{figure}

To strengthen our arguments we numerically investigate $\sigma'_{\text{reg}}(\omega)$ and the 
normalized integrated $I(\omega)$, obtained with exact (full) diagonalization where 
$\delta$-peaks are binned in windows $\delta\omega = 0.0001$. In Fig.~\ref{SI} we present 
characteristic results on $\sigma'_{\text{reg}}(\omega)$ and $I(\omega)$ for 
anisotropies $\Delta = 0.5,1$ and different magnetizations $s \sim 0,1/4$. As clearly visible, 
the regular part $\sigma'_{\text{reg}}(\omega \to 0)$ starts with zero, consistent with 
Eq.~(\ref{regu}). However, due to the small system sizes reachable by ED, it is hard to 
differentiate the predicted $\omega^2$ dependence in Eq.~(\ref{regu}) from some more general power 
law of $\sigma'_{\text{reg}}(\omega) \propto \omega^\eta$ with $\eta >0$.

Let us comment on the above results and the form of the memory function $M(\omega) = M(q=0,\omega)$, 
as deduced from the relation Eq.~(\ref{to}). If one would insert a general form 
\begin{equation}
\gamma(\omega)= \gamma_0 + \alpha \omega^r  \label{gamq0}
\end{equation}
into Eq.~(\ref{to}) and assume a nonsingular behavior of $M'(\omega)$, it is evident that the 
dissipationless component and $\sigma'_{\text{reg}}(\omega \to 0) \to 0$ are only consistent for 
$\gamma_0=0$ and $r>2$. In particular, for reproducing the form (\ref{regu}) we need $r=4$. Such a 
result clearly puts strong restrictions to proper analytical approaches to the dynamical response of 
integrable systems such as the AHM in the ``conducting'' regime.  Still some caution is 
welcome in the interpretation of our result since both analytical argument as well as numerical results 
should be firm against finite-size scaling. In particular, the relation (\ref{lc}) requires the uncorrelated
eigenvalues of $\epsilon_n$ and $j^E_n$ beyond the finite size $\omega>1/L$ (some our preliminary
results confirm this conjecture) while certain numerical results also reveal low-frequency finite-size 
anomalies \cite{herbrych11} although not for presented examples.

\section{Conclusions}
Our results confirmed the essential difference of the finite-$T$ hydrodynamic behavior between
integrable and generic nonintegrable quantum many-body systems. For the example of the dynamical 
spin susceptibility $\chi(q,\omega)$ and spin relaxation function $\Phi''(q,\omega)$ in the 1D 
generalized AHM we showed that a term breaking integrability, in our case non-zero next-neighbor 
repulsion $\Delta_2 > 0$, induces a ``normal'' diffusive behavior in the hydrodynamics, i.e.~small 
$(q,\omega)$ regime. This is well reflected in the memory-function analysis where the spin-current 
decay rate $\gamma(q,\omega)$ as the central quantity is effectively constant, 
i.e.~$\gamma(q,\omega) \sim \gamma_0$ in a range of small $(q,\omega)$. Results are in addition 
well captured within the perturbation-theory approach starting from the XX noninteracting-fermion 
model.

The integrable model with $\Delta_2=0$ we investigated in the ``conducting'' (gapless) regime, i.e.~for $s=0$, 
$\Delta<1$ and $s \neq 0$, in this way avoiding the anomalous diffusion as established in the 
``Mott-insulating'' state at $s=0, \Delta>1$.\cite{steinigeweg2012} From the existence of 
dissipationless spin transport and finite stiffness $D(T>0)>0$ it is evident that the current 
decay rate should vanish in the limit $(q,\omega) \to 0$, i.e.~$\gamma(q \to 0,\omega \to 0) = 
0$. A nontrivial result of our analysis is, however, that in this respect the universality of the 
noninteracting XX model is followed throughout the whole gapless regime revealing $\gamma(q, 
\omega=0) \propto |q|$ at $T>0$. While the similarity to the XX model occurs in the low-$\omega$ 
regime, this is not the case at larger $\omega > \omega_c = c q$. In the latter 
regime $\gamma(q,\omega>\omega_c ) = 0$ for the XX model while $\Delta>0$ induces 
$\gamma(q,\omega) \sim \gamma_1>0 $ but only weakly $(q,\omega)$-dependent. In fact, results in this 
$\omega>\omega_c$ regime again resemble the perturbation-treatment analysis, which at least 
within the lowest order  clearly fails to  capture the low-frequency dynamics.

Nontrivial are also conclusions for the uniform case $q=0$ where the results for 
$\gamma(\omega)=\gamma(q=0,\omega)$ are extracted from the uniform conductivity $\sigma'(\omega)$. 
The only consistent possibility with the dissipationless $\delta(\omega)$ component and vanishing 
regular part $\sigma'(\omega \to 0) =0$ is $\gamma(\omega \to 0) \propto \omega^r$ with $r>2$, 
e.g.~for $r=4$ the analytical argument is correct. This again indicates that in an integrable 
system the current scattering is ineffective at low $\omega$ at any $T$ in spite of even strong 
fermion interactions at $\Delta >0$. 

Although we analyzed only spin-density hydrodynamics, one can easily speculate on the behavior of 
other transport properties, in particular the energy density in the context of heat diffusion. Since 
$J^E$ is a conserved quantity, the $q=0$ dynamical thermal conductivity $\kappa(\omega)$ is 
particularly simple with only a dissipationless part and the corresponding thermal-current memory 
function $N(\omega)=0$ at any $T \geq 0$. The extension to finite-$q$ energy-density response 
$\chi_E(q,\omega)$ is not straightforward, but from the analogy to spin hydrodynamics and the 
noninteracting case one can firmly predict that the decay rate should vanish again as 
$N''(q,\omega=0) \propto |q|$. In a similar way one can possibly speculate on the $T>0$ 
hydrodynamics of other 1D integrable models such as the 1D Hubbard model.

Experimentally, the most relevant model is the isotropic $\Delta=1$ 1D Heisenberg model realized in 
several novel materials \cite{hess07}. But in the absence of external magnetic field $s=0$ case it is also the 
most controversial one. Our results in Fig.~\ref{Q} both for high $T \gg 1$ as well as for lower 
$T=0.7$ are not conclusive due to the restricted system sizes. Nevertheless, together with the 
assumption $\gamma(q\to 0, 0)=0$ (anomalous diffusion) one could expect the dependence $\gamma(q,0) 
\propto q^\alpha$ with $0<\alpha <1$. Since the effective (momentum-dependent) diffusion coefficient
${\cal D} \propto 1/\gamma(q,\omega=0)$, one could conclude that the scaling with the system size $L = 2 
\pi/q$ should follow ${\cal D} \propto L^{\alpha}$. Indeed at high temperatures such a scaling and 
divergence of ${\cal D}$ has been recently found in nonequilibrium bath scenarios with $\alpha \sim 
0.5$. \cite{znidaric11}

\acknowledgements This research was supported by the RTN-LOTHERM project and the Slovenian Agency 
grant No. P1-0044.

\end{document}